\documentclass{elsart}

\usepackage{epsfig}
\usepackage{float}
\usepackage{graphicx}    
\usepackage{dcolumn}     
\usepackage{bm}          
\usepackage{subfigure}   
\usepackage{citesort} 

\graphicspath{{Figures/}}

\begin{document}


\begin{frontmatter}
  \title{Optimal Investment Horizons for Stocks and Markets}
  \author{A.Johansen\corauthref{auth1}}
  \ead{anders-johansen@get2net.dk}
  \author{I.Simonsen\corauthref{auth2}\corauthref{auth3}}
  \ead{Ingve.Simonsen@phys.ntnu.no}
  \author{M.H.Jensen\corauthref{auth4}} \ead{mhjensen@nbi.dk}
  \address[auth1]{Teglg\aa rdsvej 119, DK-3050 Humleb\ae k, Denmark}
  \address[auth2]{Department of Physics, NTNU, NO-7491 Trondheim,
    Norway} 
  \address[auth3]{The Nordic Institute of Theoretical Physics --- NORDITA,
    Blegdamsvej 17, DK-2100 Copenhagen {\O}, Denmark}
  \address[auth4]{Niels Bohr Institute, Blegdamsvej 17,
    DK-2100 Copenhagen {\O}, Denmark}



 \begin{abstract}
   The inverse statistics is the distribution of waiting times needed
   to achieve a predefined level of return obtained from (detrended)
   historic asset prices~\cite{optihori,gainloss}.
   Such a distribution typically goes through a
   maximum at a time coined the {\em optimal investment horizon},
   $\tau^*_\rho$, which defines the most likely waiting time for
   obtaining a given return $\rho$. By considering equal positive and
   negative levels of return, we reported in~\cite{gainloss} on a
   quantitative gain/loss asymmetry most pronounced for short
   horizons. In the present paper, the inverse statistics for 2/3 of
   the individual stocks presently in the DJIA is investigated. We
   show that this gain/loss asymmetry established for the DJIA
   surprisingly is {\em not} present in the time series of the
   individual stocks nor their average. This observation points
   towards some kind of collective movement of the stocks of the index
   (synchronization). 
 \end{abstract}

\date{today}
\maketitle


\end{frontmatter}



What drives prices? This question has been studied for centuries with
quantitative theories dating back at least to Bachelier~\cite{Bachelier}, 
who proposed the random walk hypothesis for price trajectories. In order 
to qualify different pricing models 
{\it etc.}, the financial industry has performed many statistical studies 
establishing a number of stylized facts as well as benchmarking for the 
performance of various financial instruments with respect to investment 
returns and in its complement, risk taking. Due to this focus on returns 
and risk, most financial studies essentially amount to measuring two-point 
correlations in one way or another, most commonly by studying the distribution 
of returns calculated over some pre-defined fixed time period 
$\Delta t$~\cite{Book:Bouchaud-2000}. 

In recent publications~\cite{optihori,gainloss}, the present authors
have proposed to ``invert'' the standard return-distribution problem
and instead study the probability distribution of waiting times needed
to reach a {\em fixed level} of return $\rho$ for the first time.
This is in the literature known as the {\em first passage
  time}-problem and the solution is known analytically for a Brownian
motion as the inverse gamma (Levy) distribution $p(t) =
\left|a\right|\exp(-a^2/t)\ \sqrt{\pi}t^{3/2}$, (with $a\propto
\rho$), 
where one for large (waiting) times recovers the well-known first
return probability for a random walk $p(t) \sim t^{-3/2}$.  

Historical financial time series such as the DJIA, SP500 and Nasdaq
possess an often close to exponential positive drift over the time
scales of decades due to the overall growth of the economy modulated
with times of recession, wars {\it etc.}  Hence, one can obviously not
compare directly the empirical probability distribution for positive
and negative levels of return.  As the focus of the present paper will
be on such a comparison, we must ideally eliminate the effect of this
drift. We have chosen to remove the drift based on the use of
wavelets, which has the advantages of being non-parametric and hence
does not depend on certain economic assumptions.  This technique has
been described in detail elsewhere~\cite{optihori} and for the present
purpose, it suffices to say that this wavelet technique enables a
separation of the original time series $s(t)$ into a short scale
(detrended) time series $\tilde{s}(t)$ and a (long time-scale) drift
term $d(t)$ so that $s(t)=\tilde{s}(t)+d(t)$. Based on $\tilde{s}(t)$
for some historical time period of the DJIA, the empirical investment
horizon distributions, $p(\tau_\rho)$, needed to obtain a pre-defined
return level $\rho$ {\it for the first time} can easily be calculated
for different $\rho$'s. As $\tilde{s}(t)$ is stationary over time
scales beyond that of the applied wavelet (for a time larger than say
1000 days) it is straightforward to compare positive and negative
levels of return.

\begin{figure}[t]
  \centering
  \subfigure[DJIA (1896.5--2001.7)]{
    \includegraphics*[width=0.45\textwidth,height=0.45\textwidth]{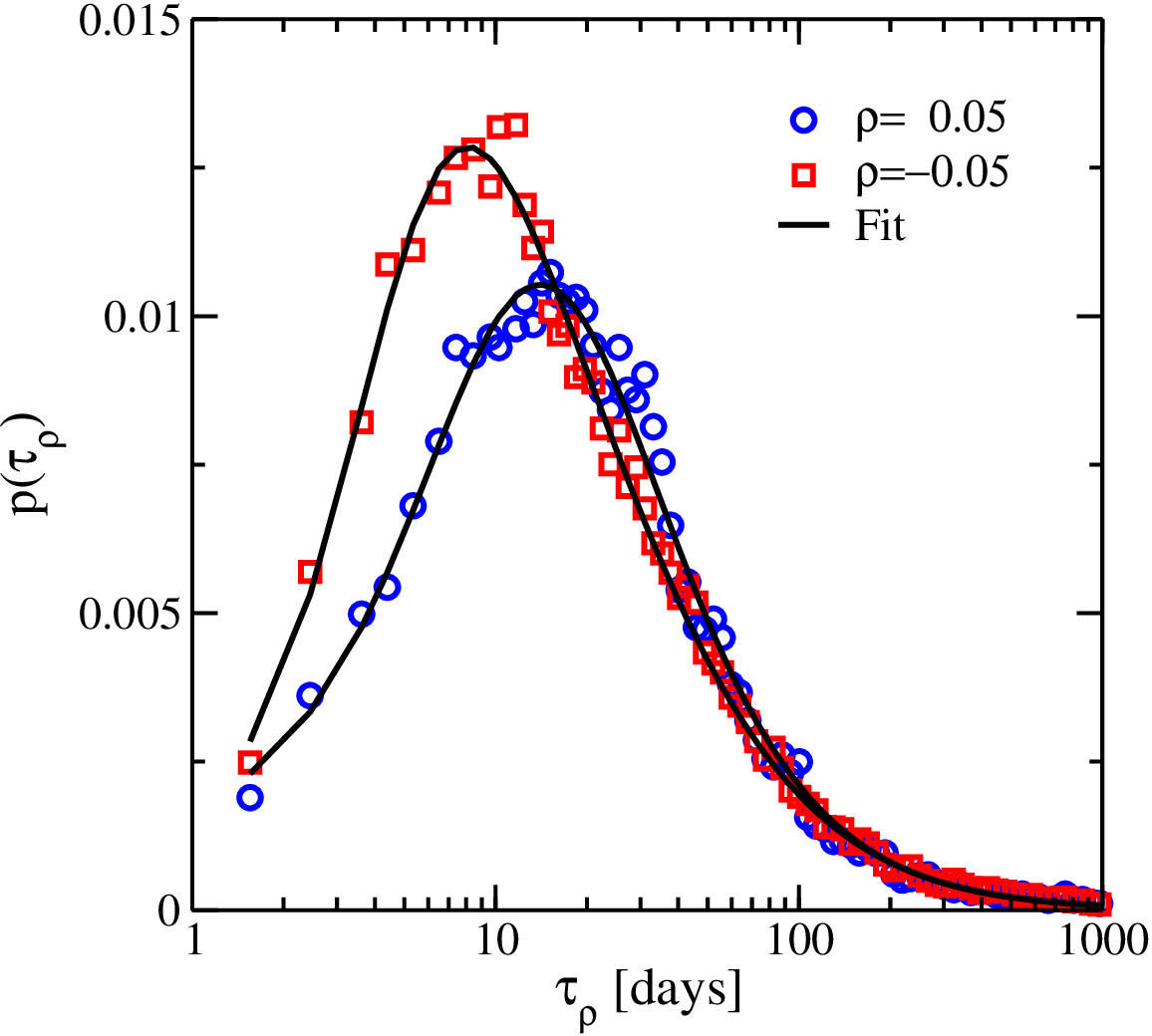} }\quad
  \subfigure[Boeing Airways (1962.1--1999.8)]{
    \includegraphics*[width=0.45\textwidth,height=0.45\textwidth]{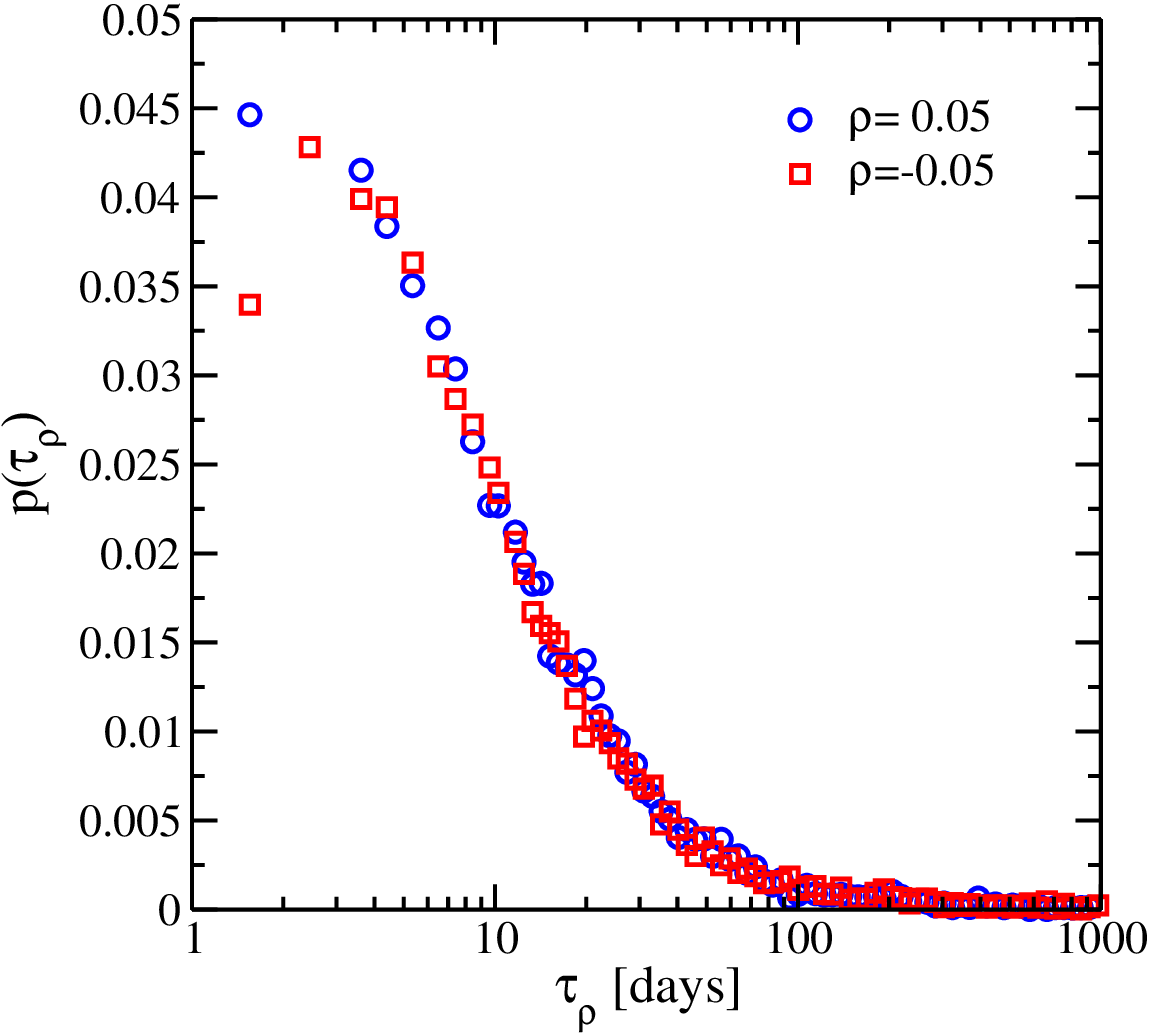} } \\*[0.5cm]
   \subfigure[General Electric (1970.0--1999.8)]{
    \includegraphics*[width=0.45\textwidth,height=0.45\textwidth]{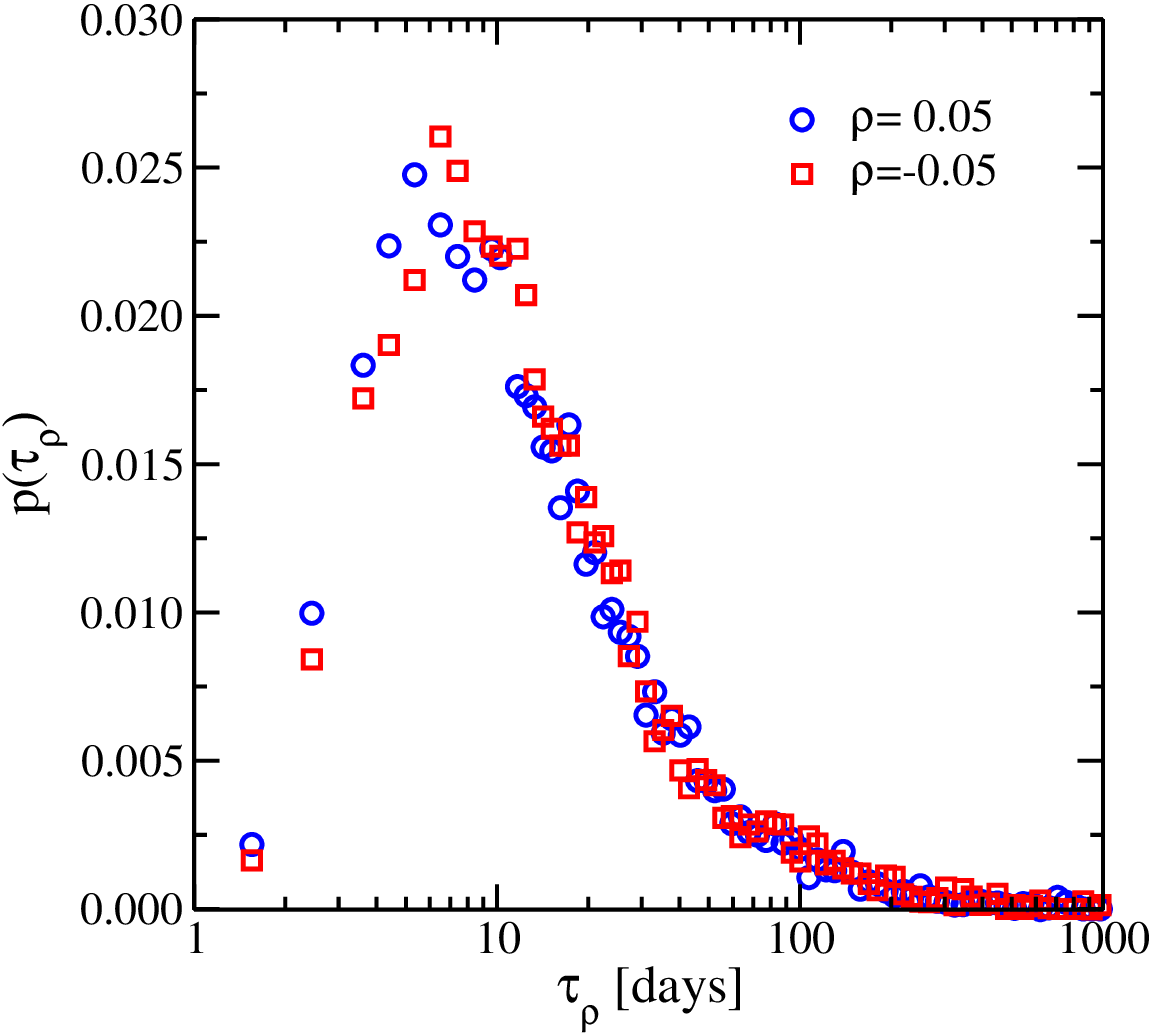} } 
\quad
   \subfigure[Stock averaged gain and loss distributions]{
  \includegraphics*[width=0.45\textwidth,height=0.45\textwidth]{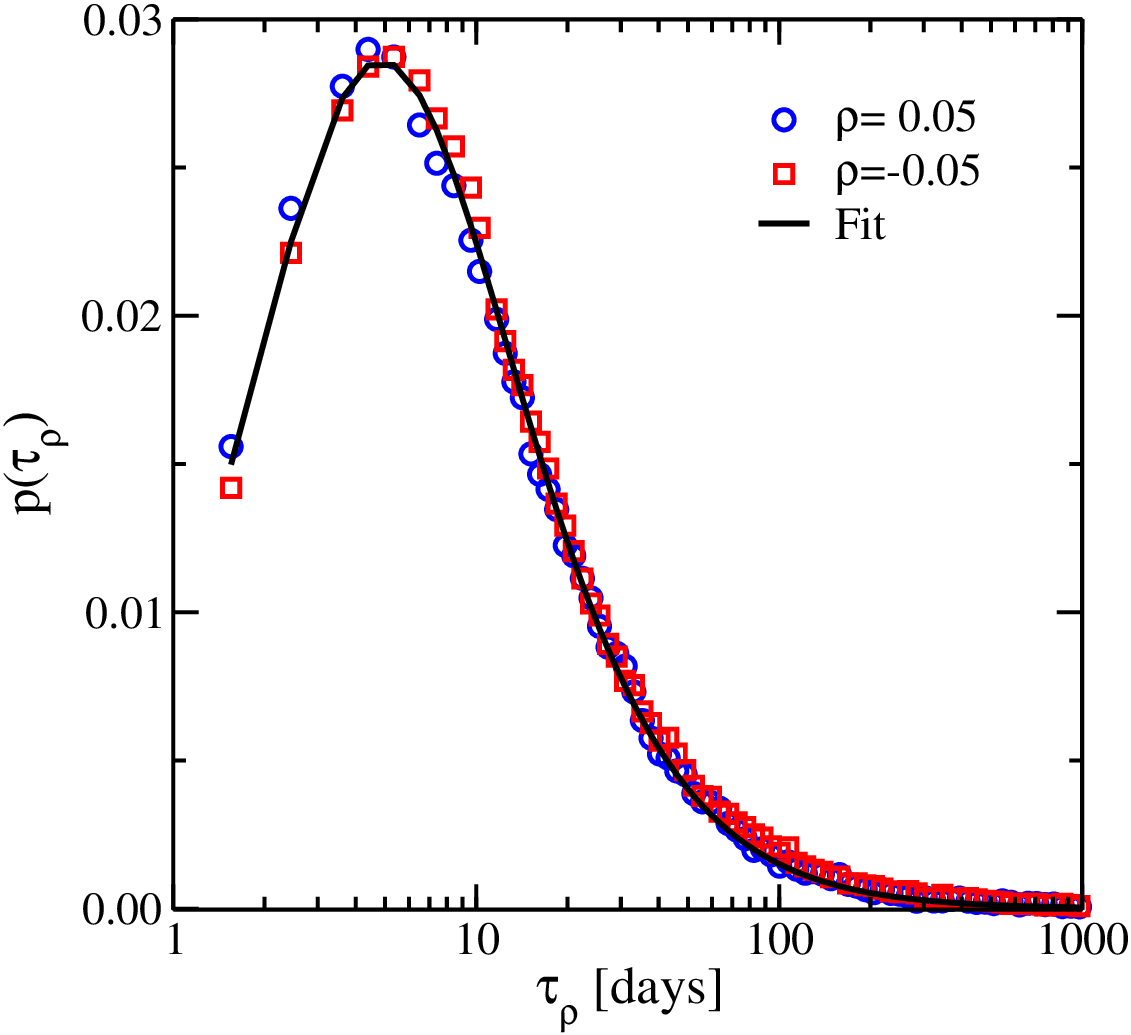} }
 
 \caption{\label{figures}Inverse statistics for detrended closing prices (open
    symbols) of the DJIA for the time periods indicated. For all cases
    the return levels used were $\left|\rho\right|=0.05$. The solid
    lines represent the best fits of Eq.~((\protect\ref{fit-func}) to
    the empirical data with the parameters indicated below; (a) DJIA
    (1896.5--2001.7): $\alpha \approx 0.51$, $\beta \approx 5.23$,
    $\nu \approx 0.68$ and $t_0 \approx 0.42$ (loss distribution);
    $\alpha \approx 0.51$, $\beta \approx 4.53$, $\nu \approx 2.13$
    and $t_0 \approx 10.1$ (gain distribution); (b) Same as (a), but for 
    Boeing Airways; (c) Same as (a), but for General Electric; (d) Stock 
    averaged gain and loss distribution for the companies listed in table 
    \protect\ref{complist}. The fit is Eq.~(\protect\ref{fit-func}) with values
    $\alpha \approx 0.60$, $\beta \approx 3.24$, $\nu \approx 0.94$ and $t_0 
    \approx 1.09$. Note that the tail exponent $\alpha+1$ is $0.1$ above the 
    ``random walk value'' of $3/2$. }

\end{figure}

As the empirical logarithmic stock price process is known not to be
Brownian, we have suggested to use a generalized (shifted) inverse
gamma distribution
\begin{eqnarray}
    \label{fit-func}
    p(t) &=&
    \frac{\nu}{\Gamma\left(\frac{\alpha}{\nu}\right)}\,
    \frac{\left|\beta\right|^{2\alpha}}{(t+t_0)^{\alpha+1} }
    \exp\left\{
          -\left(\frac{\beta^2}{t+t_0}\right)^{\nu} 
        \,\right\},
\end{eqnarray} 
where the reasons behind $t_0$ are purely technical and depends on
short-scale drift due to the fact that we are using the daily close.
The results so far have been very encouraging with respect to
excellent parametrization of the empirical probability distributions
for three major stock markets, namely DJIA, SP500 and Nasdaq; cf.
Fig.~\ref{figures}(a), for a $\rho = \pm 5\%$ example using the DJIA
and ref.~\cite{qfsubmit} for the others. The choice of $\rho=\pm 0.05$
is such that it is sufficiently large to be above the ``noise level'',
quantified by the historical volatility and sufficiently
small to occur quite frequently in order to obtain reasonable
statistics.  For all three indices, the tail-exponent $\alpha+1$
of the distributions parameterized by Eq.~(\ref{fit-func}) are
indistinguishable from the ``random walk value'' of $3/2$, which is
not very surprising. What {\it is} both surprising and very
interesting is that these three major US stock markets (DJIA, SP500
and Nasdaq) exhibit a very distinct gain/loss asymmetry, {\it i.e.},
the distributions are not invariant to a change of sign in the return
$\rho$. Furthermore, this gain/loss asymmetry quantified by the
optimal investment horizon defined as the peak position of the
distributions has for at least the DJIA a surprisingly simple
asymptotically power law like relationship with the return level
$\rho$, see \cite{gainloss} for more details.


In order to further investigate the origin of the gain/loss asymmetry
in DJIA, we simply compare the gain and loss distributions of the DJIA
with the corresponding distributions for a single stocks in the DJIA
as well as their average. An obvious problem with this approach is
that the stocks in the DJIA changes with time and hence an exact
correspondence between the DJIA and the single stocks in the DJIA is
impossible to obtain if one at the same time wants good statistics.
This is the trade-off, where we have put the emphasis on good
statistics.  The 21 company stocks analyzed and presently in the DJIA
(by the change of April 2004) are listed in Table~\ref{complist}
together with their date of entry as well as the time period of the
data set analyzed. In Figs.~\ref{figures}(c) and (c) we show the
waiting time distributions for 2 companies in the DJIA, which are
representative for the distributions obtained for all the companies
listed in Table \ref{complist}. We see that, for a return level
$\left|\rho\right|=0.05$, the value of the optimal investment horizon,
{\it i.e.} the position of the peak in the distribution, ranges from
around 2 days to around 10 days depending on the company.  More
importantly, it is clear from just looking at the figures that, within
the statistical precision of the data, the distributions are the same
for both positive and negative values of $\rho$. In order to further
quantify this invariance with respect to the sign of $\rho$, we have
averaged the (company) gain and loss distributions separately in order
to obtain an average behavior for the stocks listed in
Table~\ref{complist}.  The result of this averaging process
(Fig.~\ref{figures}(d)) is nothing less that an almost perfect
agreement between the gain and loss distributions with a peak position
around 5 days for both distributions.  This means that the optimal
investment horizon for the average of these selected individual stocks
is approximately half that of the loss distribution for the DJIA and
approximately one fourth of that for the gain distribution. In other
words, it is twice as slow to move the DJIA down and four times as
slow to move the DJIA up compared to the average time to move one of
its stocks up or down.  That market losses in general are faster than
gains must also be attributed to human psychology; people are in
general much more risk adverse than risk taking.


How can we rationalize the results we have obtained? In essence, what we have 
done is to interchange the operations of averaging over the stocks in the DJIA 
and calculating the inverse statistics for the stocks of this index. Since 
the DJIA is constructed such that it covers all sectors of the economy under 
the assumption that this areas are to some degree independent, it seems quite 
reasonable to assume that a $5\%$ gain/loss in the shares of say Boeing Airways
has nothing economically fundamental to do with a corresponding gain/loss in 
the shares of the Coca-Cola Company {\it especially} since the data are
detrended. Since the two operations do not even approximately
commute, this means that significant inter-stock correlations must exist 
even for a rather modest return level $\rho = 0.05$. 


There are several possible scenarios which may explain the observed
behavior, but they all amount to more or less the same thing. A
down/up-turn in the DJIA may be initiated by a down/up-turn in some
particular stock in some particular economical sector. This is
followed by a down/up-turn in economically related stocks and so forth. 
The result is a cascade, or synchronization, of consecutive down/up-turns in 
all the sectors covered by the DJIA and hence in the DJIA itself. The 
initiation of this may be some more general new piece of information, which 
is considered more crucial for one sector than others, but as argued 
for in length in~\cite{Schiller} it may also happen for no obvious reason 
what so ever. An (rational) example would be that Intel drops significantly 
due to bad quarterly earnings in turn, by a cascade process, affecting the
stock price of IBM and MicroSoft and so forth. As the index, at least
from a physicist's point of view, can be compared to an external
``field'', movements in the index due to a single or a few stocks can
rapidly spread through most or all sectors, {\it if} psychology in
general and specifically feed-back loops are important. The observed 
asymmetry then means that the ``field'' is not isotropic.





\newpage

%

\begin{table}
\centering
\begin{tabular}{|l||ll|} 
 \hline
 Company  & Entering date  &  Data period \\
 \hline \hline
Alcoa$^\star$           &  Apr 22, 1959   & 1962.1--1999.8  \\
American Express Company& Aug 30, 1982   & 1977.2--1999.8 \\
ATT$^\dagger$           &  Mar 14, 1939   & 1984.1--1999.8 \\
Boeing Airways          &  Jul 08, 1986   & 1962.1--1999.8 \\
Citicorp$^\bullet$      &  Mar 17, 1997   & 1977.0--1999.8 \\
Coca-Cola Company       &  Mar 12, 1987   & 1970.0--1999.8\\
DuPont                  &  Nov 20, 1935   & 1962.1--1999.8\\
Exxon \& Mobil$^\circ$  & Oct 01, 1928    & 1970.0--1999.8  \\
General Electric        &  Nov 07, 1907   & 1970.0--1999.8\\
General Motors          &  Mar 16, 1915   & 1970.0--1999.8\\
Goodyear                &  July 18 1930   & 1970.1--1999.8\\
Hewlett \& Packard      &  Mar 17, 1997   & 1977.0--1999.8\\
IBM                     &  Jun 29, 1979   & 1962.0--1999.8\\ 
Intel                   &  Nov 01, 1999   & 1986.5--1999.8\\
International Paper     &  Jul 03, 1956   & 1970.1--1999.8\\
Eastman Kodak Company   &  Jul 18, 1930   & 1962.0--1999.8\\
McDonald's Cooperation  &  Oct 30, 1985   & 1970.1--1999.8\\
Merck \& Company        &  Jun 29, 1979   & 1970.0--1999.8\\
Procter \& Gamble       &  May 26, 1932   & 1970.0--1999.8\\
The Walt Disney Co.     &  May 06, 1991   & 1962.0--1999.8\\
Wall Mart               &  Mar 17, 1997   & 1972.7--1999.8\\
\hline 
\end{tabular}
\caption{\label{complist} List of the ($21$) DJIA stocks analyzed in
  this work (about $70\%$ of the total number). Furthermore, their date 
  of entry into the DJIA are shown, and the time period covered by the 
  analyzed data set. All of these companies are also naturally part of SP500 
  with General Electric as the  most heavily weighted stock.$^\star$Former 
  Aluminum Corporation of America. $^\dagger$Former American Tel. \& Tel. 
  Truncated due to anti-trust case in 1984.  $^\bullet$Former Travelers Group. 
  $^\circ$Former Standard Oil.}
\end{table}

%

\newpage

\begin{thebibliography}{99}
\bibitem{optihori} I. Simonsen, M. H. Jensen and A. Johansen, 
Eur. Phys. J. 27 (2002) 583. 


\bibitem{gainloss} M. H. Jensen, A. Johansen and I. Simonsen, 
Physica A 234 (2003) 338. 

\bibitem{Bachelier} L. Bachelier, Th\'eorie de la Sp\'eculation", 1900, 
Annales de l'Ecole normale superiure.

\bibitem{Book:Bouchaud-2000}
J.-P. Bouchaud and M.\  Potters,
{\it Theory of Financial Risks: From Statistical 
Physics to Risk Management}
(Cambridge University Press, Cambridge, 2000).

\bibitem{qfsubmit}
A.Johansen, I.Simonsen and M.H.Jensen, 
Inverse Statistics for Stocks and Markets, preprint submitted to Quantitative
Finance, see also http://xyz.lanl.gov/physics/0511091


\bibitem{Schiller} R.\ J.\ Schiller, {\it Irrational Exuberance}, 
{Princeton University Press, 2000}



\end{thebibliography}
\end{document}